\documentclass{aastex631}

\usepackage{amsmath}
\usepackage{amsfonts}
\usepackage{amssymb}
\usepackage{placeins}
\usepackage{hyperref}
\usepackage{multirow}

\usepackage{xcolor}

\begin{document}

\newcommand{\pzvi}[1]{#1}

\title{Probing the Turbulent Corona and Heliosphere \\Using Radio Spectral Imaging Observation during the Solar Conjunction of Crab Nebula}

\author[0000-0001-6855-5799]{Peijin Zhang}
\affiliation{Center for Solar-Terrestrial Research, New Jersey Institute of Technology, Newark, NJ 07102, USA}
\affiliation{Cooperative Programs for the Advancement of Earth System Science, University Corporation for Atmospheric Research, Boulder, CO, USA}
\author[0000-0002-2325-5298]{Surajit Mondal}
\affiliation{Center for Solar-Terrestrial Research, New Jersey Institute of Technology, Newark, NJ 07102, USA}
\author[0000-0002-0660-3350]{Bin Chen}
\affiliation{Center for Solar-Terrestrial Research, New Jersey Institute of Technology, Newark, NJ 07102, USA}
\author[0000-0003-2872-2614]{Sijie Yu}
\affiliation{Center for Solar-Terrestrial Research, New Jersey Institute of Technology, Newark, NJ 07102, USA}
\author{Dale Gary}
\affiliation{Center for Solar-Terrestrial Research, New Jersey Institute of Technology, Newark, NJ 07102, USA}
\author{Marin M. Anderson}
\affiliation{Owens Valley Radio Observatory, California Institute of Technology, Big Pine, CA 93513, USA}
\affiliation{Jet Propulsion Laboratory, California Institute of Technology, Pasadena, CA 91011, USA}
\author{Judd D. Bowman}
\affiliation{School of Earth and Space Exploration, Arizona State University, Tempe, AZ 85287, USA}
\author{Ruby Byrne}
\affiliation{Cahill Center for Astronomy and Astrophysics, California Institute of Technology, Pasadena, CA 91125, USA}
\affiliation{Owens Valley Radio Observatory, California Institute of Technology, Big Pine, CA 93513, USA}
\author{Morgan Catha}
\affiliation{Owens Valley Radio Observatory, California Institute of Technology, Big Pine, CA 93513, USA}
\author[0000-0002-1810-6706]{Xingyao Chen}
\affiliation{Center for Solar-Terrestrial Research, New Jersey Institute of Technology, Newark, NJ 07102, USA}
\author[0000-0001-7754-0804]{Sherry Chhabra}
\affiliation{Center for Solar-Terrestrial Research, New Jersey Institute of Technology, Newark, NJ 07102, USA}
\affiliation{George Mason University, Fairfax, VA 22030, USA}
\author{Larry D'Addario}
\affiliation{Cahill Center for Astronomy and Astrophysics, California Institute of Technology, Pasadena, CA 91125, USA}
\affiliation{Owens Valley Radio Observatory, California Institute of Technology, Big Pine, CA 93513, USA}
\author{Ivey Davis}
\affiliation{Cahill Center for Astronomy and Astrophysics, California Institute of Technology, Pasadena, CA 91125, USA}
\affiliation{Owens Valley Radio Observatory, California Institute of Technology, Big Pine, CA 93513, USA}
\author{Jayce Dowell}
\affiliation{University of New Mexico, Albuquerque, NM 87131, USA}
\author{Katherine Elder}
\affiliation{School of Earth and Space Exploration, Arizona State University, Tempe, AZ 85287, USA}
\author{Gregg Hallinan}
\affiliation{Cahill Center for Astronomy and Astrophysics, California Institute of Technology, Pasadena, CA 91125, USA}
\affiliation{Owens Valley Radio Observatory, California Institute of Technology, Big Pine, CA 93513, USA}
\author{Charlie Harnach}
\affiliation{Owens Valley Radio Observatory, California Institute of Technology, Big Pine, CA 93513, USA}
\author{Greg Hellbourg}
\affiliation{Cahill Center for Astronomy and Astrophysics, California Institute of Technology, Pasadena, CA 91125, USA}
\affiliation{Owens Valley Radio Observatory, California Institute of Technology, Big Pine, CA 93513, USA}
\author{Jack Hickish}
\affiliation{Real-Time Radio Systems Ltd, Bournemouth, Dorset BH6 3LU, UK}
\author{Rick Hobbs}
\affiliation{Owens Valley Radio Observatory, California Institute of Technology, Big Pine, CA 93513, USA}
\author{David Hodge}
\affiliation{Cahill Center for Astronomy and Astrophysics, California Institute of Technology, Pasadena, CA 91125, USA}
\author{Mark Hodges}
\affiliation{Owens Valley Radio Observatory, California Institute of Technology, Big Pine, CA 93513, USA}
\author{Yuping Huang}
\affiliation{Cahill Center for Astronomy and Astrophysics, California Institute of Technology, Pasadena, CA 91125, USA}
\affiliation{Owens Valley Radio Observatory, California Institute of Technology, Big Pine, CA 93513, USA}
\author{Andrea Isella}
\affiliation{Department of Physics and Astronomy, Rice University, Houston, TX 77005, USA}
\author{Daniel C. Jacobs}
\affiliation{School of Earth and Space Exploration, Arizona State University, Tempe, AZ 85287, USA}
\author{Ghislain Kemby}
\affiliation{Owens Valley Radio Observatory, California Institute of Technology, Big Pine, CA 93513, USA}
\author{John T. Klinefelter}
\affiliation{Owens Valley Radio Observatory, California Institute of Technology, Big Pine, CA 93513, USA}
\author{Matthew Kolopanis}
\affiliation{School of Earth and Space Exploration, Arizona State University, Tempe, AZ 85287, USA}
\author{Nikita Kosogorov}
\affiliation{Cahill Center for Astronomy and Astrophysics, California Institute of Technology, Pasadena, CA 91125, USA}
\affiliation{Owens Valley Radio Observatory, California Institute of Technology, Big Pine, CA 93513, USA}
\author{James Lamb}
\affiliation{Owens Valley Radio Observatory, California Institute of Technology, Big Pine, CA 93513, USA}
\author[0000-0002-4119-9963]{Casey J Law}
\affiliation{Cahill Center for Astronomy and Astrophysics, California Institute of Technology, Pasadena, CA 91125, USA}
\affiliation{Owens Valley Radio Observatory, California Institute of Technology, Big Pine, CA 93513, USA}
\author{Nivedita Mahesh}
\affiliation{Cahill Center for Astronomy and Astrophysics, California Institute of Technology, Pasadena, CA 91125, USA}
\affiliation{Owens Valley Radio Observatory, California Institute of Technology, Big Pine, CA 93513, USA}
\author{Brian O'Donnell}
\affiliation{Center for Solar-Terrestrial Research, New Jersey Institute of Technology, Newark, NJ 07102, USA}
\author{Kathryn Plant}
\affiliation{Owens Valley Radio Observatory, California Institute of Technology, Big Pine, CA 93513, USA}
\affiliation{Jet Propulsion Laboratory, California Institute of Technology, Pasadena, CA 91011, USA}
\author{Corey Posner}
\affiliation{Owens Valley Radio Observatory, California Institute of Technology, Big Pine, CA 93513, USA}
\author{Travis Powell}
\affiliation{Owens Valley Radio Observatory, California Institute of Technology, Big Pine, CA 93513, USA}
\author{Vinand Prayag}
\affiliation{Owens Valley Radio Observatory, California Institute of Technology, Big Pine, CA 93513, USA}
\author{Andres Rizo}
\affiliation{Owens Valley Radio Observatory, California Institute of Technology, Big Pine, CA 93513, USA}
\author{Andrew Romero-Wolf}
\affiliation{Jet Propulsion Laboratory, California Institute of Technology, Pasadena, CA 91011, USA}
\author{Jun Shi}
\affiliation{Cahill Center for Astronomy and Astrophysics, California Institute of Technology, Pasadena, CA 91125, USA}
\author{Greg Taylor}
\affiliation{University of New Mexico, Albuquerque, NM 87131, USA}
\author{Jordan Trim}
\affiliation{Owens Valley Radio Observatory, California Institute of Technology, Big Pine, CA 93513, USA}
\author{Mike Virgin}
\affiliation{Owens Valley Radio Observatory, California Institute of Technology, Big Pine, CA 93513, USA}
\author{Akshatha Vydula}
\affiliation{School of Earth and Space Exploration, Arizona State University, Tempe, AZ 85287, USA}
\author{Sandy Weinreb}
\affiliation{Cahill Center for Astronomy and Astrophysics, California Institute of Technology, Pasadena, CA 91125, USA}
\author{David Woody}
\affiliation{Owens Valley Radio Observatory, California Institute of Technology, Big Pine, CA 93513, USA}




\begin{abstract}
Measuring plasma parameters in the upper solar corona and inner heliosphere is challenging because of the region's weakly emissive nature and inaccessibility for most in situ observations. Radio imaging of broadened and distorted background astronomical radio sources during solar conjunction can provide unique constraints for the coronal material along the line of sight. In this study, we present radio spectral imaging observations of the Crab Nebula (Tau A) from June 9 to June 22, 2024 when it was near the Sun with a projected heliocentric distance of 5 to 27 solar radii, using the Owens Valley Radio Observatory's Long Wavelength Array (OVRO-LWA) at multiple frequencies in the 30--80 MHz range. The imaging data reveal frequency-dependent broadening and distortion effects caused by anisotropic wave propagation through the turbulent solar corona at different distances. We analyze the brightness, size, and anisotropy of the broadened images. 
Our results provide detailed observations showing that the eccentricity of the unresolved source increases as the line of sight approaches the Sun, suggesting a higher anisotropic ratio of the plasma turbulence closer to the Sun. In addition, the major axis of the elongated source is consistently oriented in the direction perpendicular to the radial direction, suggesting that the turbulence-induced scattering effect is more pronounced in the direction transverse to the coronal magnetic field. 
Lastly, when the source undergoes large-scale refraction as the line of sight passes through a streamer, the apparent source exhibits substructures at lower frequencies.
This study demonstrates that observations of celestial radio sources with lines of sight near the Sun provide a promising method for measuring turbulence parameters in the inner heliosphere.

\end{abstract}


\section{Introduction} \label{sec:intro}

The upper corona and inner heliosphere region in $\sim$5--30 solar radii play crucial roles in connecting the solar activity and space weather phenomena. Understanding the plasma parameters in these regions is essential for solar and space weather studies. However, diagnosing the plasma in such a region presents significant challenges due to the region's weak-emissive nature. 
White light coronagraphs such as the Large Angle Spectroscopic Coronagraph (LASCO) \citep{Brueckner1995LASCO} can provide observational constraints for large-scale plasma density distribution and reveal dynamic structures such as coronal mass ejections and streamers. 
In-situ observation \citep{coles1991solar} can measure the local plasma properties at a given point.
The plasma properties at intermediate scales (or ``mesoscale'') are critical to study as it is a key factor in the development and evolution of solar wind turbulence, and can influence the transport of energetic particles.
However, it remains difficult to measure mesoscale properties such as density fluctuations and their anisotropic degree.

In this context, radio astronomical techniques have emerged as powerful tools for probing the outer corona and inner heliosphere regions.
By observing the apparent radio source from unresolved background celestial sources as the emission traverses the turbulent solar corona during its solar conjunction, one can infer critical plasma characteristics through the effects of radio wave propagation through the coronal plasma along the line of sight. Refraction and scattering effects can cause temporal variations in the observed radio flux, as well as spatial variations of the apparent radio source in shape and position. The study of the temporal variation of radio sources, referred to as the interplanetary scintillation (IPS) method (pioneered by \citealt{Hewish1964Natur}), is well established, which provides insights into the density and turbulence of the interplanetary medium.

With spatially resolved radio observation from radio telescope arrays, it is possible to observe the angular broadening and orientation of the broadening to derive the scattering properties of the plasma density fluctuation and its anisotropic degree. Early studies with the Very Large Array \citep{ Armstrong1990VLA, grall1997observations} have demonstrated that the anisotropic ratio of the source can be derived by forward fitting the visibility with free parameters, including the axial ratio and the orientation angle $\theta$ at the frequencies of 1.5 and 4.8~GHz. 
The Crab Nebula (Tau-A), a bright and relatively compact source in low frequencies ($<$200MHz), is particularly suitable for the study of angular broadening.
\pzvi{The angular broadening phenomena were first reported by \cite{hewish1955irregular,hewish1958scattering}, who measured the size of Tau A with an interferometer, at wavelengths of 3.7~m and 7.9~m}. \citet{ramesh2001low} reported an increase in the amplitude of the angular broadening of Tau-A when a CME crossed the line of sight, indicating there is enhanced density fluctuation within the CME.
\cite{raja2016amplitude, raja2017turbulent, raja2019dissipation} discussed the density turbulence and proton heating rate based on the observation of Tau-A's solar conjunction in 2012-2013.

Despite the long-term interest in this area, there remains a lack of low-frequency wideband observations that explore the frequency-dependent behavior of anisotropic angular broadening. 
As shown by \citet{kontar2023anisotropic}, multiple factors, including the electron density, the amplitude of density fluctuations, and the degree of anisotropy in the fluctuations, govern angular broadening. These parameters are strongly coupled, making the interpretation of angular broadening complex. Observations at a single frequency can provide limited quantitative diagnostics, while multi-frequency imaging is essential for more comprehensive and robust constraints. By comparing the parametric simulation and the multi-frequency observation, one may place better constraints on the multiple factors above and as well as providing a diagnosis on the plasma property.
Additionally, high-sensitivity and high-dynamic-range observations are needed to investigate features associated with solar approaches at low frequencies.

Tau-A at low frequencies ($<100MHz$) is bright and stable, making it a widely used calibration source. Its flux spectrum has been extensively measured and is well known \citep{de2020cassiopeia}. 
However, when Tau-A's line of sight \pzvi{passes through the solar corona}, the plasma with a higher and presumably more fluctuating electron density in the solar vicinity causes increased scattering and absorption of the radio waves \citep{kontar2019anisotropic,kontar2023anisotropic}. Consequently, the observed total flux density is reduced. As the scattering and absorption effect is stronger for lower frequencies, the effect of reduced flux is more pronounced. The observation of the flux density of the source and its frequency dependency can thus be used to estimate the absorption parameters, as well as the distribution and fluctuations of density in the outer corona and inner heliosphere.

In addition to probing the turbulence properties of the general solar corona, it is also of interest to detect such properties within specific coronal structures, such as streamers. 
Solar streamers are fundamental components of the coronal architecture \citep{Brueckner1995LASCO}.
However, measuring the extended density distribution of streamers, particularly in the outer corona, remains challenging. Previous efforts, such as \citet{decraemer2019three}, have combined white-light coronagraph observations with forward modeling to infer the density distribution of streamers.
In this context, the density variation of the streamer can be embedded into the observed source when its line of sight passes through the streamer. This potential capability offers a novel viewing angle to investigate streamer structures. 
Despite its potential, such use of radio source substructures to study large-scale coronal features has not been systematically explored before. 



In this study, we present new radio spectral imaging observations of Tau A during its solar transit from June 9 to June 22, using the Owens Valley Radio Observatory's Long Wavelength Array (OVRO-LWA) in the frequency range of 30--80 MHz. By analyzing the frequency-dependent brightness, size, and anisotropy of the source broadening as the line of sight to Tau A approaches the Sun at various heliocentric distances, we provide a more comprehensive understanding of the plasma properties in the upper coronal and inner heliospheric distances. Taking advantage of the high-dynamic-range, high-sensitivity images made by OVRO-LWA, we found that when the source approaches the Sun, it presents arc shapes and has substructures when the source is on the verge of attenuation towards the invisible. 
This paper is arranged as follows: in Section 2, we present the observation and data processing, including all the measurements. In section 3, we discuss scattering based on the observational results. Then we summarize and discuss the results in Section 4.

\FloatBarrier
\section{Observation}

\begin{figure}
    \centering
    \includegraphics[width=0.69\linewidth]{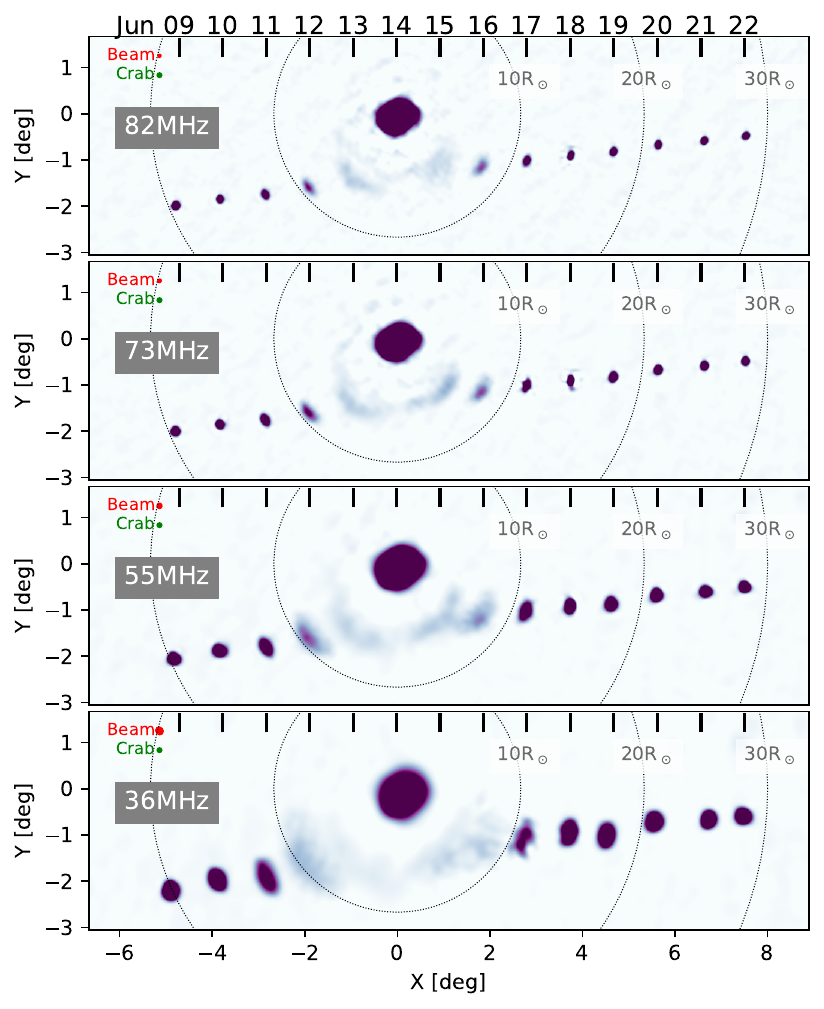}
    \caption{Time-lapse overlap of the Crab Nebula during the solar transit from 2024 June 9 to June 22. Each panel shows the radio image observed at 20:30:05 UT$\,\pm\,$5s (near local noon at OVRO) on the respective days. The corresponding date is indicated by the red bars on top of each panel. The image of the Sun shown in all panels is from 2024 June 12th. The FWHM size of the synthesized beam at each frequency and the intrinsic size of the Crab Nebula are marked by red and green circles, respectively.}
    \label{fig:main}
\end{figure}

The observation was performed with the OVRO-LWA, a low-frequency radio interferometer composed of 352 crossed broadband dipole antennas.
The OVRO-LWA has all-sky imaging capability as the correlator does full-cross-correlation for the 352 antennas.
It operates in the frequency range of 13.4 MHz to 86.9 MHz, with a 26 kHz channel width, and 192 channels per band (5 MHz).
The standard imaging mode has a cadence and integration time of 10 s.

We collected 3 minutes of data every day from 9th June 2024 to 22nd June 2024 at 20:30 UT (near noon of local time of the observatory), in the frequency range of 23--87 MHz. The lowest bands were not used due to strong radio frequency interference (RFI) and ionosphere activities below 23 MHz.
The recorded visibility data, as the sampling in the Fourier space (or ``uv'' space), was processed in the following steps: 
\begin{itemize}
    \item [1.] Apply bandpass solutions from the nighttime calibrator observation.
    \item [2.] Direction-independent self-calibration: two iterations of phase-only calibration and three iterations of amplitude-phase calibration. 
    \item [3.] Imaging step to convert visibility into the flux density distribution image by gridding and CLEAN process to deconvolve the point-spread-function (PSF) from the resulting image. 
    \item [4.] Finally, cutting out the region of interest and performing source fitting to get the source brightness, size, and shape. Prepare for further analysis.
\end{itemize}

 The data processing utilizes the open-sourced package \texttt{ovro-lwa-solar}\footnote{\texttt{ovro-lwa-solar} \href{https://github.com/ovro-eovsa/ovro-lwa-solar}{https://github.com/ovro-eovsa/ovro-lwa-solar}}, which builds upon the radio interferometry data analysis package \texttt{CASA}\footnote{\texttt{CASA} \href{https://casadocs.readthedocs.io/en/stable/}{https://casadocs.readthedocs.io/en/stable/}},  radio synthesis imaging software \texttt{WSClean}\footnote{\texttt{WSClean} \href{https://gitlab.com/aroffringa/wsclean}{https://gitlab.com/aroffringa/wsclean} \citep{offringa2014wsclean}}, and common astronomy software package \texttt{AstroPy}.
An overview of the observation and processed images of Tau A transiting the Sun is shown in Figure \ref{fig:main} as a time-lapse stitched image.
From Figure \ref{fig:main}, one can see that the source size and brightness of Tau A clearly change when the source approaches the Sun.
There is stronger angular broadening when the source is closer to the Sun. The source gradually becomes fainter when closer to the Sun, and the source is almost invisible during its closest approach on June 14th at all frequencies.
In the following subsections, we present a detailed analysis of the source's shape and brightness during the solar transit period.


\subsection{Angular broadening and orientation}

As shown in Figure \ref{fig:main}, the observed source shape can be approximated by a 2D Gaussian function.
Thus, we adopt a 2D Gaussian fit to measure the shape and orientation of the source.
The brightness distribution of the source can be expressed as:
    $$ I(x,y)  =  I_0 \exp\left( -\frac{x'^2}{2 \sigma_x^2} - \frac{y'^2}{2 \sigma_y^2} \right)$$
    where
    $$ x' = (x-x_0) \cos\theta_T -(y-y_0)\sin\theta_T$$
    $$y' = (x-x_0) \sin\theta_T -(y-y_0)\cos\theta_T$$
where $x$ and $y$ are along the right ascension and declination direction, respectively.
$I_0$ is the peak value of the distribution, $x_0$, $y_0$ is the center of the model source. $\sigma_x$, $\sigma_y$ is the standard deviation of the 2D Gaussian in major and minor axes, respectively, and $\theta_T$ is the tilting angle relative to north, measured eastwards. 

\begin{figure}
    \centering
    \includegraphics[width=0.99\linewidth]{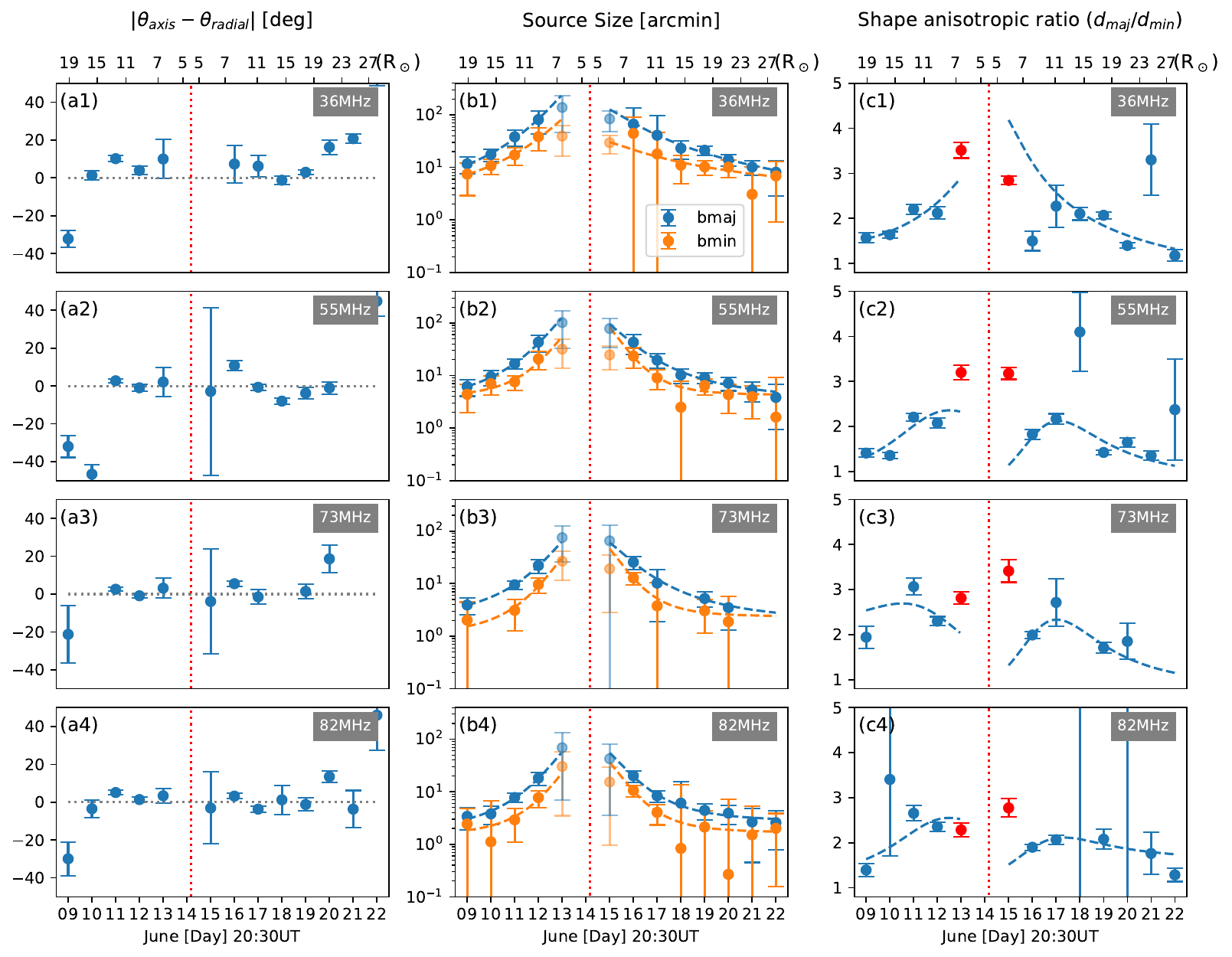}
    \caption{Deconvolved 2D elliptical Gaussian source shape statistics of Crab (Tau-A) during the transit, the bottom x-axis is the date of the observation, the observation time is 20:30 UT of each day, the top x-axis shows the distance of the Crab to the Sun as solar radius.
    Panels \textbf{(a*)} show the angular distance of the minor axis and radial direction of different frequencies. Panels \textbf{(b*)} present the source size measured as Major (blue) and Minor (orange) axes. Panels \textbf{(c*)} is the anisotropic ratio, measured as the ratio of the major and minor axes.
    The statistics and fitted results are also shown in Table \ref{table:main}.
    The red points in Column \textbf{(c)} are from the angular Gaussian fitting method.
    \pzvi{We note that the data point of June 13 and 15 is marked as half-transparent in (b*) and red in (c*) because it is measured by fitting an angular 2D Gaussian, detailed in Section 2.2.}
    }
    \label{fig:sourceshape}
\end{figure}

Based on a 2D Gaussian model, \texttt{casa}'s \texttt{imfit} tool\footnote{\texttt{imfit}:  \href{https://casadocs.readthedocs.io/en/latest/api/tt/casatasks.analysis.imfit.html}{https://casadocs.readthedocs.io/en/latest/api/tt/casatasks.analysis.imfit.html}} is used to derive the deconvolved source shape $I_{deconv}$ where:
$$
I_{conv} = I_{deconv} \ast \rm{beam}
$$
The deconvolved source shape provides a more accurate estimate of the intrinsic source properties by minimizing the influence of the synthesized beam on the observed image.
Figure \ref{fig:sourceshape} shows the fitted result of the deconvolved 2D Gaussian model. Column (a) shows the orientation, defined as the angular difference between the source's minor axis ($\theta_{axis}$) and the radial direction connecting the Crab to the center of the Sun ($\theta_{radial}$). Column (b) displays the full width at half maximum (FWHM) of the source. Column (c) shows the axial ratio, defined by the ratio of major to minor axis. Summary statistics and fitting results are also shown in Table \ref{table:main}. \pzvi{The intrinsic size of Tau A, as reported by \citet{de2020cassiopeia}, is 7.9 arcminutes with a bright compact central core at 50MHz. This size is indicated by a green circle in the upper-left corner of Figure~\ref{fig:main}. At higher frequencies ($>55$~MHz), the source size is comparable to the synthesized beam, and at lower frequencies ($<55$~MHz), it is smaller than the beam. Therefore, Tau A is treated as an unresolved source throughout this work.}

From the shape measurements shown in Figure \ref{fig:sourceshape} and Table \ref{table:main}, we can have the following results:
\begin{itemize}
    \item [1.] When the source displays a significant eccentricity, the orientation (as shown in panel (a) in Figure \ref{fig:sourceshape}) of the source's minor axis aligns very well with the radial direction of the Sun (the average deviation angle is within 8 degrees). This alignment indicates that turbulence-induced scattering is stronger in the tangential direction, perpendicular to the solar radial direction.
    \item [2.] The source size (as shown in panel (b) in Figure \ref{fig:sourceshape}) has a trend of getting larger when closer to the Sun. The trend is similar in all frequencies.
    This indicates there is stronger scattering in the line-of-sight for the source when it is closer to the Sun.
    \item [3.] The average axial ratio ($\sigma_{\rm major}/\sigma_{\rm minor}$) across different frequencies ranges from 1.7 to 2.5.
    The ratio increases as the line of sight of the source approaches the Sun, suggesting that background density fluctuations are more anisotropic closer to the Sun than farther away.
\end{itemize}
We need to note that, the observation of June 14th is not presented because the source is not detected, the observation of June 13th and 15th is not included in the statistics in Table \ref{table:main} as it is measured with a different method from the other shape measurements, detailed in the next subsection.

\begin{table}[h!]
\centering
\begin{tabular}{|c|c|c|c|c|c|}
\hline
\multirow{2}{*}{Frequency} & \multirow{2}{*}{Trail} & \multirow{2}{*}{$\theta_{axis} - \theta_{radial}$ } &  $d=b(R/R_\odot)^a+c$ &  $d=b(R/R_\odot)^a+c$ & \multirow{2}{*}{$d_{maj}/d_{min}$} \\
 & & & bmaj (a, b, c) & bmin (a, b, c) & \\
\hline
\multirow{2}{*}{36MHz} & Ingress & 3.03 $\pm$ 16.27 & -3.31, 13.2$\times 10^4$, 3.89 & -3.08, 3.15$\times 10^4$, 4.17 & 1.88 $\pm$ 0.33 \\
 & Egress & 7.91 $\pm$ 25.56 & -1.63, 0.20$\times 10^4$, 7.09 & -2.89, 1.37$\times 10^4$, 6.52 & 1.71 $\pm$ 0.67 \\
\hline
\multirow{2}{*}{55MHz} & Ingress & -3.90 $\pm$ 21.77 & -3.60, 11.6$\times 10^4$, 3.50 & -4.99, 99.9$\times 10^4$, 4.54 & 1.74 $\pm$ 0.39 \\
 & Egress & 4.56 $\pm$ 31.88 & -2.57, 0.79$\times 10^4$, 3.46 & -4.28, 14.4$\times 10^4$, 4.35 & 1.71 $\pm$ 0.87 \\
\hline
\multirow{2}{*}{73MHz} & Ingress & 1.70 $\pm$ 12.28 & -3.03, 1.60$\times 10^4$, 1.79 & -3.37, 1.38$\times 10^4$, 0.81 & 2.42 $\pm$ 0.41 \\
 & Egress & 3.36 $\pm$ 7.44 & -2.00, 0.15$\times 10^4$, 0.13 & -1.73, 0.04$\times 10^4$, 1.47 & 2.02 $\pm$ 0.38 \\
\hline
 \multirow{2}{*}{82MHz} & Ingress & 1.08 $\pm$ 13.03 & -3.22, 2.07$\times 10^4$, 1.47 & -3.61, 1.73$\times 10^4$, 1.43 & 2.19 $\pm$ 0.65 \\
 & Egress & 1.76 $\pm$ 15.81 & -2.65, 0.40$\times 10^4$, 2.29 & -2.38, 0.11$\times 10^4$, 1.02 & 1.94 $\pm$ 4.33 \\
\hline

\end{tabular}
\caption{Source shape statistic and fitting results in Figure \ref{fig:main}, first column ($\theta_{axis} - \theta_{radial}$) is the average angle difference of the minor axis of the source and the radial direction from the Sun. The second and third columns are the source size fitting results for major and minor axes, and the fourth column is the anisotropic ratio of the source. }
\label{table:main}
\end{table}


\subsection{Source deformation near solar transit}

\begin{figure}
    \centering
    \includegraphics[width=0.68\linewidth]{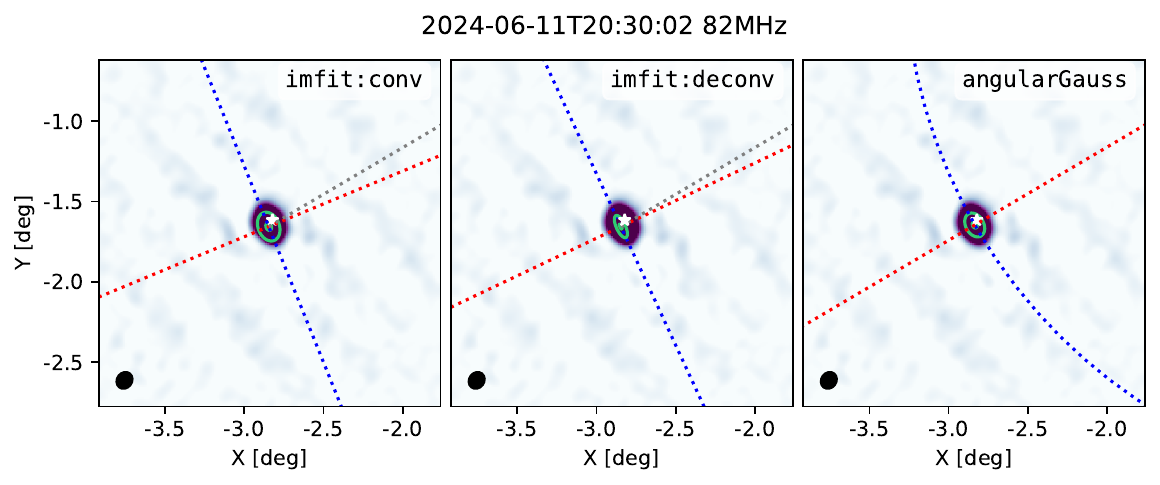}
    \includegraphics[width=0.68\linewidth]{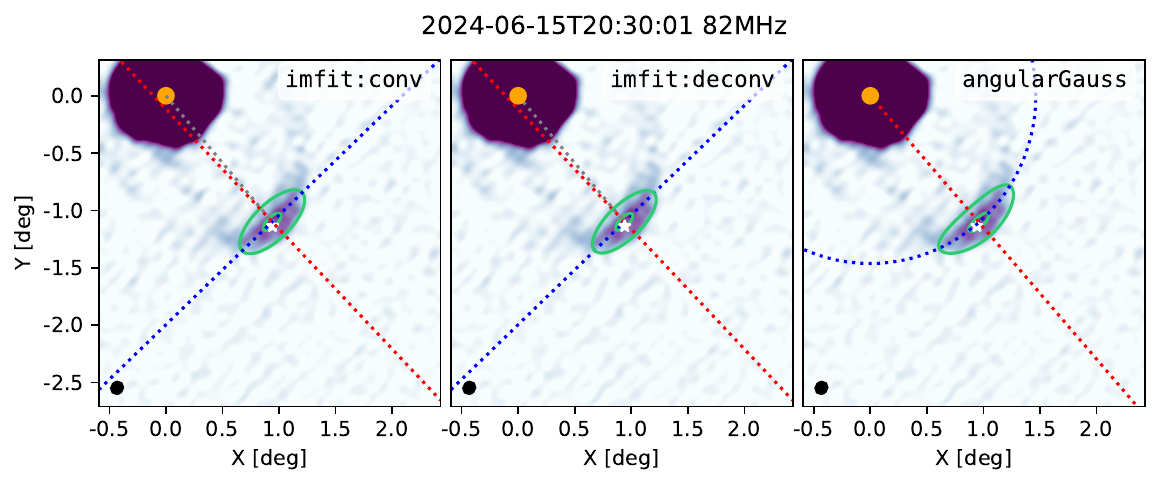}
    
    \caption{The source shape modeled as 2D elliptical Gaussian and angular Gaussian, for the Crab imaging of 82MHz on 2024 June 15th.
    The left panel is the 2D Gaussian, and the blue and red dotted line shows the direction of the major and minor axes. The right panel is the angular Gaussian fitted result, the red and blue line indicates the $\theta$ direction and $R$ radius.}
    \label{fig:angularGaussian}
\end{figure}

We found that on June 13th and 15th, when the source's line of sight is at 6 solar radii from the Sun, the source shows an arc shape instead of a simple 2D Gaussian, as shown in Figure \ref{fig:angularGaussian}.
In this case, the source shape is not well described by a 2D Gaussian model, as in the other days with larger distances.
Hence, we introduce an anisotropic Gaussian model in polar coordinates to provide a more accurate description of the source.
This Gaussian model extends in the direction of tangential angular ($\theta$) and radial ($r$) directions, rather than $x$ and $y$ as in the typical 2D Gaussian model in Cartesian coordinates. The expression is:
    $$I(x,y)  = I_0 \exp\left( - \frac{(R-R_0)^2}{2 \sigma_R^2} - \frac{(\theta-\theta_0)^2}{2 \sigma_\theta^2} \right)$$
    where,
    $$R = \sqrt{x^2+y^2}$$
    $$\theta = \arctan{(x/y)}$$
In the model, $R$, $\theta$ is the polar coordinate element, $\sigma_R$ is the rms width in the radial direction, and $\sigma_\theta$ is the rms angular width of the source measured from the source center ($R_0$, $\theta_0$). The radial and tangential directions are defined with respect to the coordinate center of the Sun obtained from the image with the method introduced in.

Figure~\ref{fig:angularGaussian} shows the comparison of the fitting methods, from which we can see that, for the source with large separation distances (e.g., June 11th), the fitting result using the anisotropic polar Gaussian model is similar to that by using the Cartesian 2D Gaussian model. As the source size at this frequency on June 11th is at a similar scale as the beam size, it is more appropriate to fit the deconvolved source based on the Cartesian 2D Gaussian model (as shown in the middle column of Figure \ref{fig:angularGaussian}). For the source on June 15th, as the source size is much larger than the beam size, as expected, the fitting of the original observed source and beam-deconvolved source yields similar results. In addition, since the source on June 15th shows an arc shape, the anisotropic polar Gaussian model achieves a better description of the source shape in this case (right panel).

\begin{figure}
    \centering
    \includegraphics[width=0.95\linewidth]{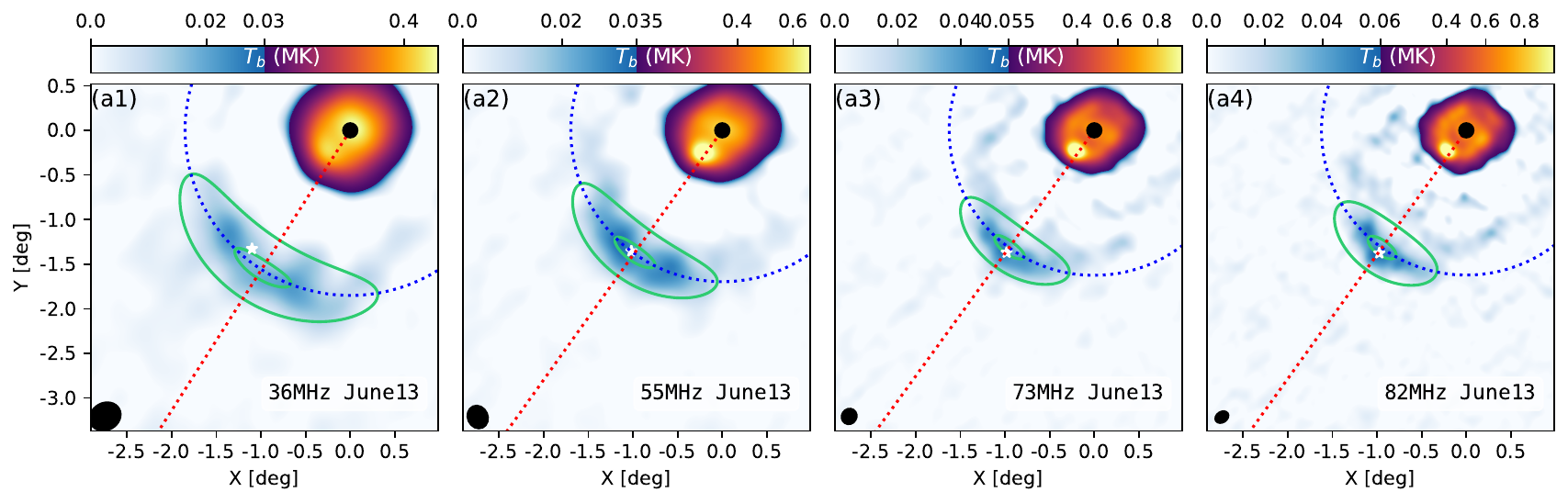}
    \includegraphics[width=0.95\linewidth]{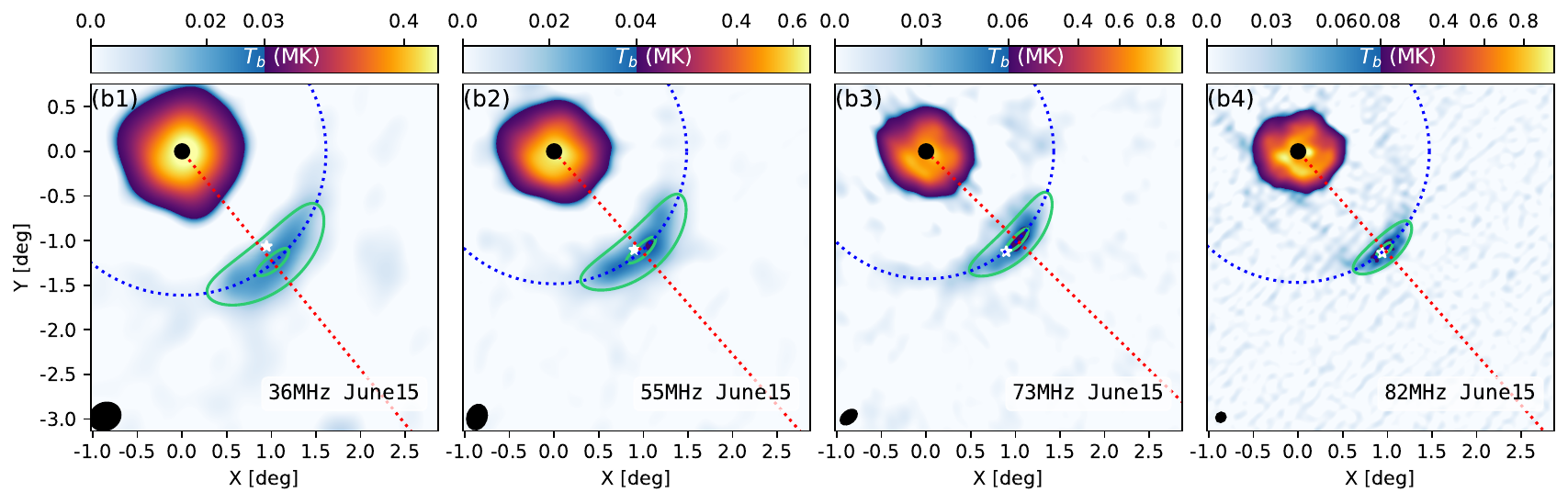}
    \caption{The imaging result of the source on June 13th and 15th, overlapped with the fitted result of Angular Gaussian Distribution as green-solid-line for the contour of 0.5 and 0.9 of the peak. The white star indicates the intrinsic coordinate of the Crab Nebula.}
    \label{fig:arcs} 
\end{figure}

\begin{figure}
    \centering
    \includegraphics[width=0.6\linewidth]{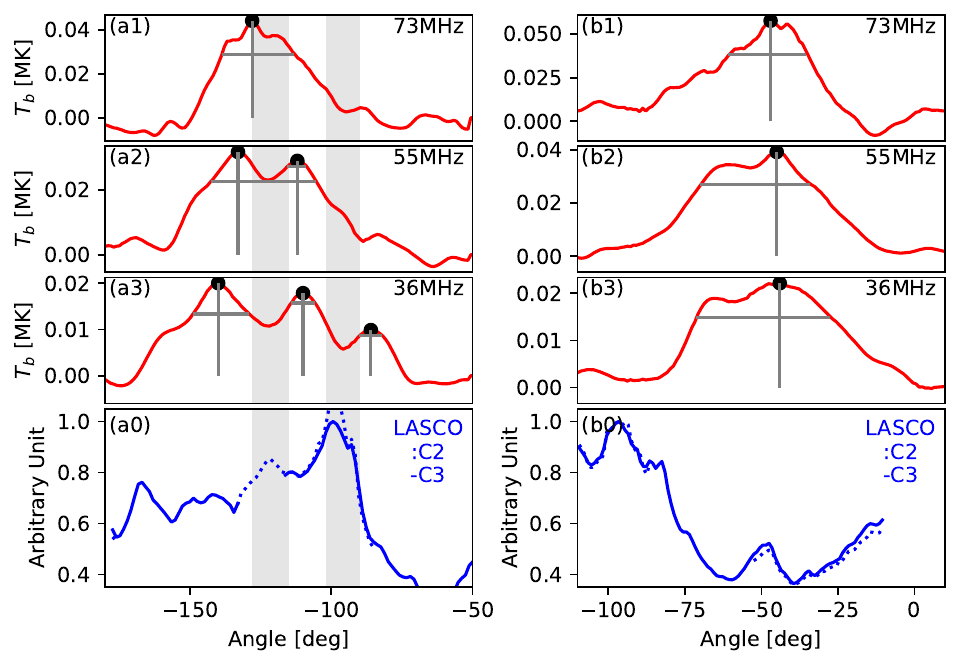}
    \includegraphics[width=0.19\linewidth]{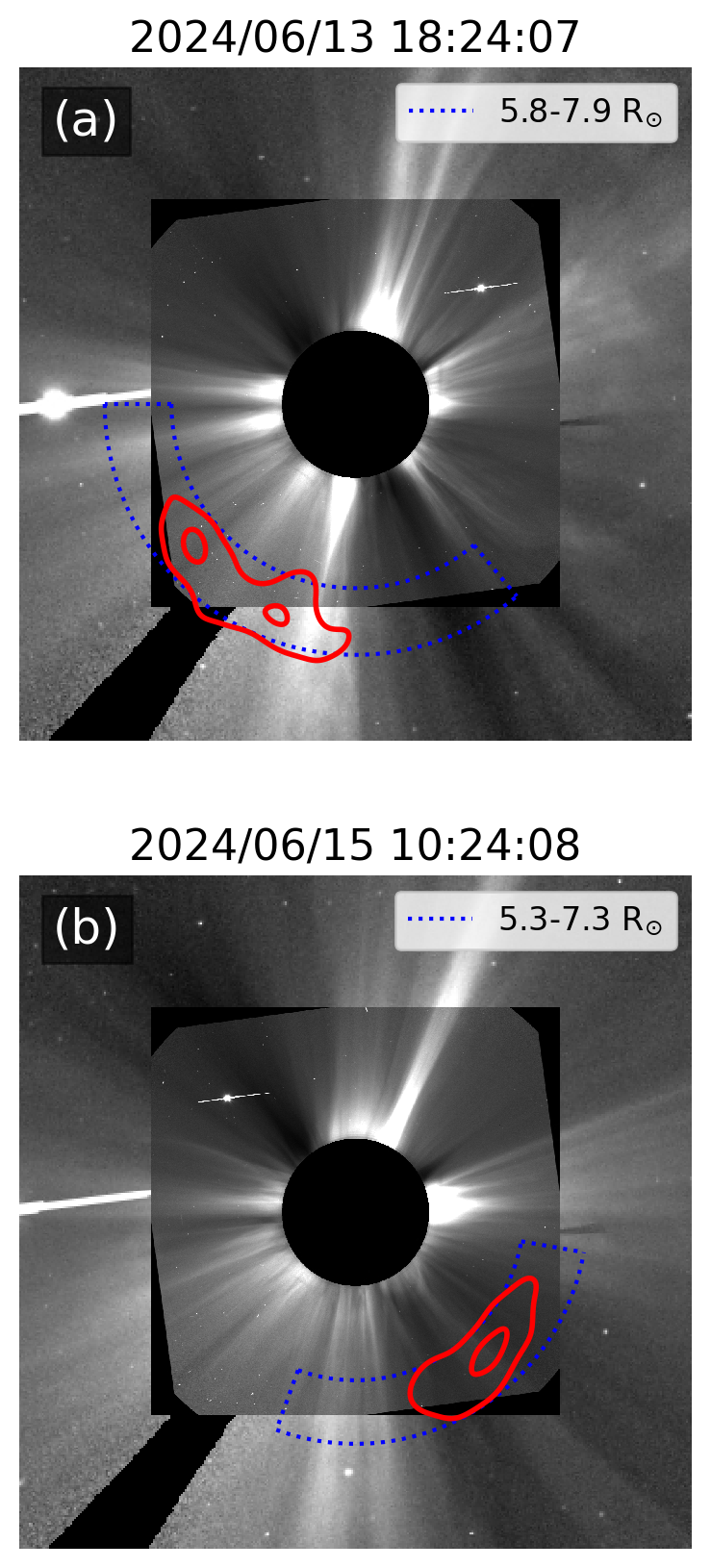}
    
    \caption{Brightness angular distribution of the Crab on June 13 and June 15. Panels (a 1-3) and (b 1-3) present the brightness temperature distribution sampled at a heliocentric distance of the Crab on different position angles (anti-clockwise from the East direction).
    The black dot marks the peak location in the distribution. 
    The blue line in panels (a0) and (b0) shows the angular distribution of white-light brightness from LASCO.
    Panels (a) and (b) present the contour of the source at 36MHz overlay on LASCO C2/C3 white light image of the corresponding nearest time point.}
    \label{fig:ang_dist}
\end{figure}

On June 13th and 15th, when Tau A was during its ingress and egress of the solar transit at small projected heliocentric distances (6.4 $R_{\odot}$ and 5.6 $R_\odot$, respectively), the source exhibits distinct substructures at lower frequency bands (36 MHz and 55 MHz). Despite this, its overall arc-shaped brightness distribution can still be characterized using the anisotropic polar Gaussian model.
The brightness distribution along the angular direction (Figure \ref{fig:ang_dist} panels (a\,1-4) and (b\,1-4)) indicates that there are more substructures in the source of June 13th (ingress) than that of June 15th (egress): on June 13th at 36 MHz, the source exhibits three prominent peaks, and at 55 MHz the angular distribution shows two well-defined peaks. 
A notable feature is the anti-correlation between the angular distributions of the white-light and radio emissions (panels (a3) and (a0)). Specifically, regions with enhanced white-light brightness—indicative of higher plasma density—correspond to regions of diminished radio flux. This suggests that the radio wave is absorbed or re-distributed by the higher-density streamers.
For the source observed on June 15th, however, the angular distribution presents much less prominent substructures. 
Also, the source size in the tangential direction on the 13th is larger than on the 15th.
According to the white light coronagraph data, a clear difference during the ingress and egress at similar heliocentric distances lies in that the source's line of sight on June 13th passes through a more pronounced streamer region compared to that on June 15th, as shown in Figures~\ref{fig:ang_dist}(a) and (b).
The observed sub-structures in the source morphologies suggest that, on top of the scattering effects induced by the overall coronal plasma, they may carry additional information about density fluctuations associated with smaller-scale coronal structures, such as streamers, at spatial scales of 0.1--1$^{\circ}$.

\subsection{Source brightness}

\begin{figure}
    \centering    
    \includegraphics[width=0.64\linewidth]{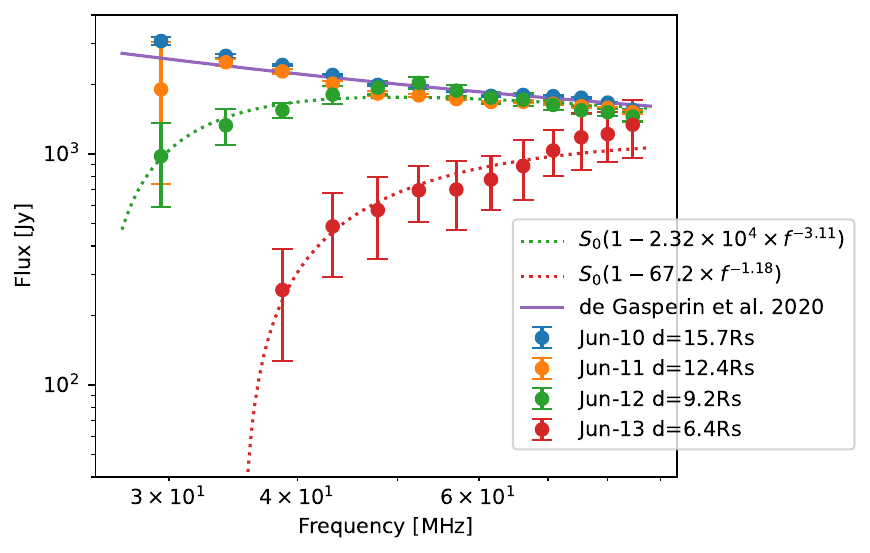}
    \caption{Flux spectrum of the Crab compared with the model from \cite{de2020cassiopeia} in the frequency range of 30-85MHz. The purple curve shows the model spectrum. The colored points with error-bars present the flux spectrum of the Crab on Jun 10,11,12,13 respectively.
    The green and red dotted lines are power-law fits for the attenuation. \pzvi{X-axis is frequency (MHz) in log-scale}.}
    \label{fig:fluxSpec3day}
\end{figure}

The total flux of the source is measured by integrating the flux density over the source area.
We measured the total flux for 13 frequency bands ranging from 23 to 87 MHz with a bandwidth of 5 MHz for each band. 
Figure \ref{fig:fluxSpec3day} shows the total flux measurement results of Tau A on June 10th, 11th, and 12th (the corresponding heliocentric distance in projection is 15.7, 12.4, and 9.2 solar radii, respectively).

The observed frequency-brightness relation can be well described by a frequency-dependent attenuation based on an empirical model of the source brightness:
$$
S_{obs}(f) =  S_0(f)\times(1- a f^b)
$$
where $S_0$ is the model flux from \cite{de2020cassiopeia}.
From the spectrum, we can see that the spectrum on June 10th, when Tau A is located a large projected heliocentric distance of 15.7 $R_{\odot}$, is well consistent with the empirical model of \cite{de2020cassiopeia}, meaning there is no noticeable influence on the measured total flux density from the heliospheric medium at this distance.
The spectrum on June 11, when Tau A approaches a slightly closer distance of 12.4 solar radii, only shows one data point deviating from the model at 28 MHz, although the deviation is comparable to the measurement uncertainties. 
In comparison, the spectra on June 12 and June 13 present a significant deviation from the model flux, especially in the low-frequency end.
In particular, on June 13th, the absorption caused attenuation of more than 10 times for observation frequencies lower than 40 MHz. 
The total flux of June 12th and June 13th can be expressed using an empirical function as:
$$
S_{06/12} = S_0( 1-2.32 \times 10^4 × f^{-3.11})
$$
$$
S_{06/13} = S_0( 1-67.2 \times f^{-1.18})
$$
The results indicate strong absorption or scattering attenuation effects for the line of sight close to the Sun, and the effect is more significant for lower frequencies.

\FloatBarrier

\section{Scattering and anisotropic ratio}

\begin{figure}
    \centering
    \includegraphics[width=0.96\linewidth]{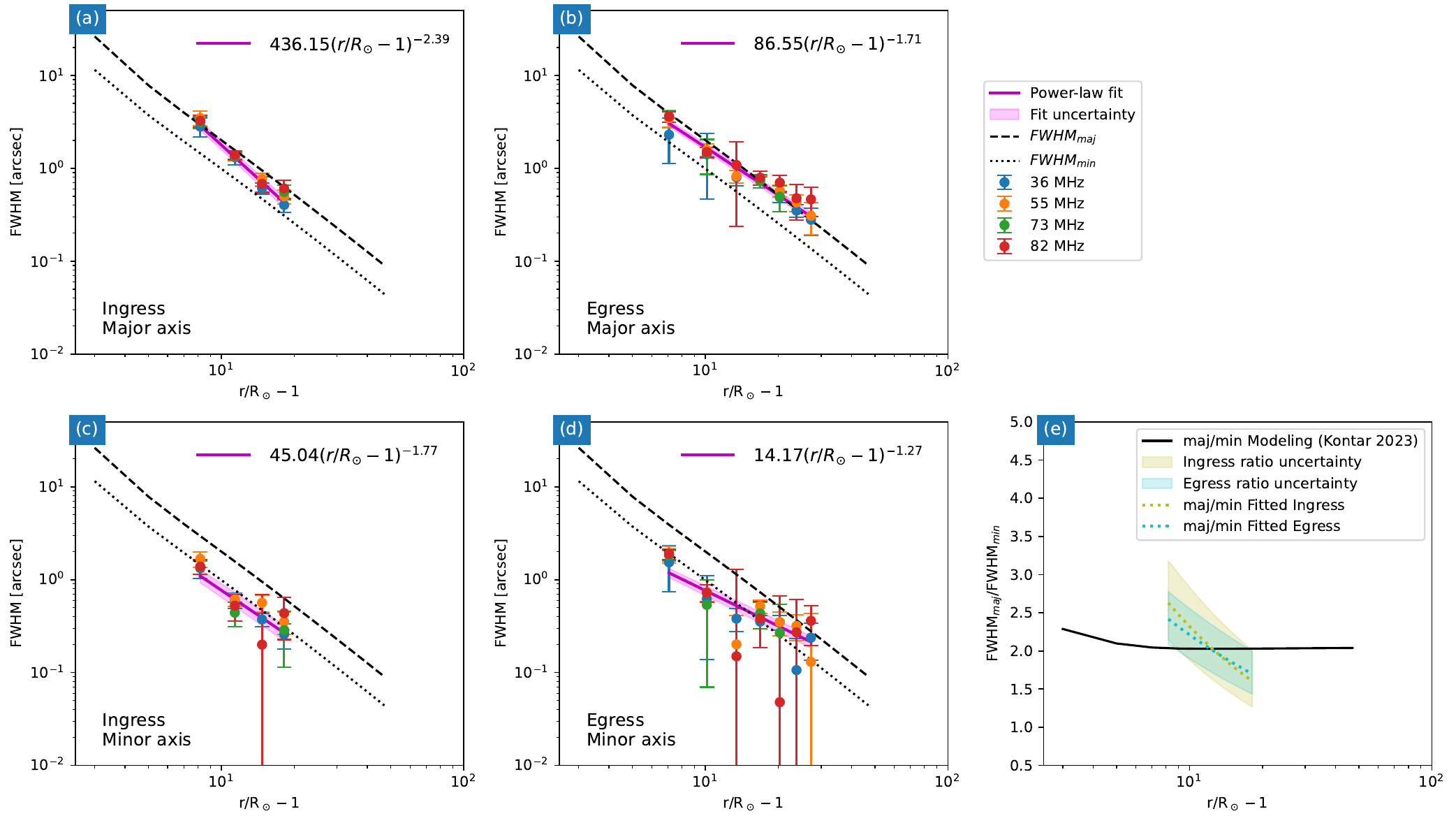}
    \caption{Comparison of the observed source size and the modeled angular broadening size from the model (both normalized to 1.5~GHz). 
    The panels (a) and (c) present the observation from Ingress, and (b) and (d) present that from Egress. The panels (a) and (b) present the measurements of the major axis, and the panels (c) and (d) present the measurements of the minor axis. 
     In each panel, the black dashed line and the black dotted line show the modeled angular size of the major and minor axes. The colored dots with error bars are from the observation in this work, and the solid magenta line is the power-law fit from the observation in the current panel. The modeled size is derived from the model in 
     \cite{kontar2023anisotropic} with parameter of anisotropic ratio of density fluctuation $\alpha=0.25$. The source size in this plot is normalized to 1.5~GHz.}
    \label{fig:scatMod}
\end{figure}

The angular broadening of a distant point source with its line-of-sight passing through the turbulent corona or inner heliosphere has been modeled by ray tracing simulations  \citep{kontar2019anisotropic,kontar2023anisotropic}.
The model can be used to emulate the scattering process along the line of sight and estimate the observed angular size of the broadened source.
There is a key factor in the simulation: the anisotropic ratio ($\alpha=q_\parallel/q_\perp$), which represents the ratio of the wavenumber of the density fluctuation in the parallel or perpendicular direction. It was reported as $\alpha$ in the range of 0.25-0.5 using various types of observations.
In this work, to compare with the observed source shape, \pzvi{we derived the expected observed source shape with numerical simulation} with the model in \citet{kontar2023anisotropic},  using an assumed value of $\alpha=0.25$. 
The model-predicted angular FWHM size, shown in Figure~\ref{fig:scatMod} as black dashed and dotted lines for the major and minor axes, respectively, has a constant source axial ratio $\sigma_{maj}/\sigma_{min}\approx2.0$ from 2 to 40 solar radii. 

From observation, we can measure the size of the major and minor axes during the ingress and egress period. The fitted results are as follows, \pzvi{we note that the source shape from June 13 to June 15 is not included in the fitting for the consideration of different fitting schemes and occultation}:
\begin{align}
\theta_{[\text{Ingress Major}]} &= (436.15 \pm 27.08) \times (r/R_\odot-1)^{(-2.39 \pm 0.02)}   \\
\theta_{[\text{Egress Major}]} &= (86.55 \pm 4.07) \times (r/R_\odot-1)^{(-1.71 \pm 0.02)} 
   \\
\theta_{[\text{Ingress Minor}]} &= (45.04 \pm 4.03) \times (r/R_\odot-1)^{(-1.77 \pm 0.03)} 
  \\
\theta_{[\text{Egress Minor}]} &= (14.17 \pm 0.95) \times (r/R_\odot-1)^{(-1.27 \pm 0.02)} 
\end{align}
By comparing the modeled angular broadening from the observation and the modeling, we can see that the major axis and minor axis sizes generally agreen with the model-predicted values from \citep{kontar2023anisotropic} with $\alpha=0.25$.

However, the observed axial ratio exhibits a systematic deviation from the model predictions: it is generally larger than the model at smaller heliocentric distances and smaller than the model at larger distances (as shown in Figure \ref{fig:scatMod}(e)). This trend suggests that the plasma density fluctuations closer to the Sun are more anisotropic than assumed in the model with a constant $\alpha = 0.25$, while those further away are less anisotropic than $\alpha = 0.25$. 
At approximately $r \sim 10\,R_\odot$, the observed source shape is consistent with the model prediction using $\alpha = 0.25$. 
Thus, our results indicate that there is a radial variation in anisotropy, with stronger anisotropy in the inner corona and weaker anisotropy in the outer corona.

\section{Discussion and Conclusion}

In this study, we conducted low-frequency radio (23--87 MHz) observations of Tau-A when its line of sight was passing close to the Sun at small angular distances from June 9 to June 22, 2024. 
We carried out a detailed analysis of angular broadening and brightness variations during this period, which are used to diagnose anisotropic scattering and absorption processes by the coronal plasma along the line of sight. 
The key findings from the observations can be summarized as follows:
\begin{itemize}

\item The observed anisotropic feature of the broadened source, in general, agrees with the predictions from the stochastic wave scattering model  \citep{kontar2023anisotropic}.
The measured value of axial ratio ($b_{maj}/b_{min}$) ranges from 1.7 to 2.5, which is consistent with the scattering model \citep{kontar2023anisotropic} predicted a fixed value of 2.0 (derived with $\alpha=0.25$) in the range of 6-40 solar radius.
The shape axial ratio of the source has a trend of increasing with the decrease of the separation angle, indicating more anisotropic density fluctuations for the coronal medium near the Sun.

\item At a small separation angle ($<$2$^{\circ}$) and lower frequency ($<$60 MHz), the angular broadening extending into a tangential direction reveals an arc shape.

\item For the first time, sub-structures within the angular broadened source were observed. These substructures appeared when the source’s line of sight intersected strong streamer regions in the corona. We found an anti-correlation between the brightness angular distribution of white-light and the substructure.

\item The brightness-frequency relation of the source was measured and established, serving as a reference for future studies on absorption and scattering attenuation.
\end{itemize}

\pzvi{The size and shape reported in this work are, in general, in good agreement with previous observations. 
The size from the early work by \cite{1963HewishMNRASsolar} was measured with the fringe pattern of the interferometer at 38 MHz, with the assumption that the brightness profile follows a Gaussian distribution as a function of baseline length. The observation showed that the size distribution as a function of the heliocentric distance could be described by a power-law function with a form of $\theta \propto(R/R_\odot)^{\beta}$, with a power-law index ranging from $-$1.30 to $-$2.24 in the range of 10--20 $R_\odot$ based on observations from 1959 to 1962. We have similar results in the range of 6--19 $R_\odot$, with a power-law index ranging from $-1.27$ to $-2.39$ for $\theta \propto(R/R_\odot-1)^{\alpha}$ for major and minor axes of ingress and egress, respectively.
The axial ratio is difficult to determine directly from interferometric fringe patterns. In previous works \citep{hewish1958scattering, Hewish1964Natur}, an assumed value of 2:1 was used. \citet{raja2017turbulent} reported an axial ratio of approximately 2 with interferometric imaging observations, with larger values observed in the presence of streamers, though no systematic variation with heliocentric distance was noted. In this work, we also measure an axial ratio of $\approx 2$, consistent with previous studies, but we additionally find a clear trend of increasing axial ratio as the line of sight approaches the Sun.}

The observed features in this study include the frequency and distance dependence of angular broadening, arc-shaped source morphology, substructures within the broadened sources, and brightness reduction near the Sun. These features are strongly linked to coronal structures across multiple spatial scales. The angular broadening itself arises from small-scale density fluctuations along the line of sight, with its shape reflecting the amplitude and anisotropic ratio of the turbulence. Meanwhile, the radial trend in broadening and brightness attenuation is related to large-scale variations in the background coronal density and fluctuation amplitude.
Meso-scale structures such as coronal streamers further modulate the scattering: when the line of sight intersects these dense, structured regions, the resulting broadened source exhibits arc-like deformation and internal substructures. These signatures carry the imprint of streamer-scale density gradients and inhomogeneities. Thus, angular broadening is a good indicator of the density variation of the coronal across different scales.
Forward modeling of radio-wave scattering through anisotropic turbulence, as demonstrated in \citet{kontar2023anisotropic}, provides a theoretical framework to interpret these observables. 
With future observations of additional compact sources near the Sun and expanded multi-frequency coverage, combined with advanced wave propagation simulations, we can place tighter constraints on key coronal parameters—including the spatial distribution of density fluctuations, streamer profiles, and the variation of anisotropy with heliocentric distance.

The anisotropic nature of coronal plasma turbulence has been extensively explored in theoretical and simulation studies \citep[e.g.,][]{higdon1984density, chen2016recent, kontar2023anisotropic}. 
A key parameter used to characterize the degree of anisotropy is $\alpha$, defined as the ratio of wavenumbers parallel and perpendicular to the anisotropy axis \citep{kontar2019anisotropic}. This parameter ranges from 0 to 1, where $\alpha = 1$ corresponds to fully isotropic turbulence, and $\alpha \gtrsim 0$ indicates strong anisotropic fluctuation with fluctuations predominantly perpendicular to the anisotropic axis.
Despite its importance in describing plasma turbulence, the anisotropy ratio $\alpha$ has remained difficult to constrain observationally. Previous studies \citep{zhang2021parametric, kontar2023anisotropic, clarkson2024magnetic} have attempted to estimate $\alpha$ by combining angular broadening, scintillation, and radio burst observations, yielding typical values in the range of 0.25–0.4. However, these works have not reported any significant spatial variation of $\alpha$ with heliocentric distance.
Here, thanks to the availability of unprecedentedly high-quality, wide-field imaging data from the newly commissioned OVRO-LWA, we provide a more detailed analysis using full source properties, including the time- and frequency-dependent measurements along both the major and minor axes of the angularly broadened source.
These new spectral imaging observations enable us to constrain the anisotropic ratio of the density fluctuation ratio: $\alpha$.
Our results demonstrate that while an average value of $\alpha \approx 0.25$ broadly describes the observed angular broadening, there is a clear systematic trend: the anisotropy becomes stronger (i.e., smaller $\alpha$) at smaller heliocentric distances. This suggests that the density turbulence in the inner corona is more strongly guided by the magnetic field, reinforcing the interpretation of magnetically controlled anisotropic turbulence in the low corona.
A further systematic observational constraint on $\alpha$ as a function of distance provides critical input for solar wind models that rely on turbulence properties. It helps the understanding of how energy is redistributed across scales and directions. Also a more precise estimation of anisotropic ratio of the solar wind can help the radio wave propagation study to have a better estimate of the source size and location based on ray tracing simulations.



Significant flux attenuation is observed as the source's line of sight nears the Sun, especially at lower frequencies ($<40$\,MHz). On June 12 and 13, at heliocentric distances of 9.2 and 6.4 solar radii, respectively, the measured flux density significantly deviates from the \citet{de2020cassiopeia} model, with attenuation exceeding $\times10$ below 40\,MHz on June 13. This strong attenuation can be attributed to enhanced scattering and absorption of the radio wave in the turbulent coronal plasma, which is stronger at lower frequencies and higher electron density regions according to \cite{kontar2019anisotropic, kontar2023anisotropic}. 
The frequency-dependent attenuation offers a potential diagnostic tool for probing plasma properties of the higher corona: by parametrically modeling the scattering and absorption deriving the source brightness attenuation, and comparing with the observed values.

This work demonstrates the potential of using angular broadening observations to probe the corona and inner heliosphere. With more sensitive radio telescopes offering high fidelity and high dynamic range imaging, we expect to resolve more sources at smaller separation angles to the Sun. This capability can help us derive key properties of the coronal structures along multiple lines of sight using different point sources, yielding a new and more comprehensive understanding of the upper coronal and inner heliospheric plasma structures. 

\section{Acknowledgements}
P. Z. acknowledges support for this research by the NASA Living with a Star Jack Eddy Postdoctoral Fellowship Program, administered by UCAR’s Cooperative Programs for the Advancement of Earth System Science (CPAESS) under award 80NSSC22M0097. The OVRO-LWA expansion project was supported by NSF under grant AST-1828784. OVRO-LWA operations for solar and space weather sciences are supported by NSF under grant AGS-2436999. 





%

\vspace{5mm}
\facilities{OVRO-LWA}


\software{astropy}



\bibliography{cite}{}
\bibliographystyle{aasjournal}



\end{document}